\documentclass[preprint2]{aastex6}
\usepackage{color}
\usepackage[titletoc]{appendix}
\usepackage[fleqn]{amsmath}
\usepackage{float}
\usepackage{threeparttable}
\usepackage{comment}
\usepackage{enumitem}
\usepackage{natbib}
\usepackage{bm}

\shortauthors{Millholland \& Laughlin}
\shorttitle{Obliquity Tides in WASP-12b}

\begin{document} 

\title{Obliquity Tides May Drive WASP-12b's Rapid Orbital Decay} 
\author{Sarah Millholland$^{1,2}$ and Gregory Laughlin$^{1}$}
\affil{$^1$ Department of Astronomy, Yale University, New Haven, CT 06511;  \href{mailto:sarah.millholland@yale.edu}{sarah.millholland@yale.edu}} 
\altaffiliation{$^2$ NSF Graduate Research Fellow}

\begin{abstract}
Recent analyses have revealed a mystery. The orbital period of the highly inflated hot Jupiter, WASP-12b, is decreasing rapidly. The rate of inspiral, however, is too fast to be explained by either eccentricity tides or equilibrium stellar tides. While dynamical stellar tides are possible, they require a subgiant structure for the star, whereas stellar models point toward a main sequence host. Here, we show that these hitherto irreconcilable observations might be explained by planetary obliquity tides if planet b's spin vector is trapped in a high-obliquity state maintained by a secular spin-orbit resonance with an unseen exterior perturbing planet. We derive constraints on the obliquity ($\epsilon\gtrsim50^{\circ}$), reduced tidal quality factor ($Q^{\prime}\sim10^{6}-10^{7}$), and perturbing planet parameters ($M_{2}\sim10-20M_{\oplus}$, $a_2\lesssim0.04\,{\rm AU}$) required to generate the observed orbital decay. Direct N-body simulations that include tidal and spin dynamics reinforce the plausibility of the scenario. 
Furthermore, we show that the resonance could have been captured when planet b's obliquity was small, making the proposed sequence of events easy to explain. 
The hypothetical perturbing planet is within the limits of current radial velocity constraints on the system yet is also detectable. If it exists, it could provide evidence in favor of the \textit{in situ} formation hypothesis for hot Jupiters.

\end{abstract}

\section{Introduction}

With their ponderous masses, torrid orbits, and swollen radii, hot Jupiters are prone to the influence of tidal forces \citep{1996ApJ...470.1187R,2004ApJ...610..477O}.  The slow action of tides delineates the structural and dissipative properties of these alien worlds and also points back to their origins. (See \citealt{2014ARA&A..52..171O} for a review.) 
To date, however, WASP-12b is the only hot Jupiter whose tidal evolution can be observed in real time. This extremely inflated ($R_p\approx{1.9}R_{\mathrm{Jup}}$, $M_p\approx{1.4}M_{\mathrm{Jup}}$) gas giant was discovered by \cite{2009ApJ...693.1920H} orbiting a late-F main sequence star. \cite{2016A&A...588L...6M} and \cite{2017AJ....154....4P} measured the planet's $1.0914$ day transit period to be decreasing on a rapid $P/\dot{P}=-3.2$ Myr timescale. 

Apsidal precession could explain the period decrease if the planet's eccentricity is maintained by dynamical perturbations. \cite{2016A&A...588L...6M} and \cite{2017AJ....154....4P}, however, found that this scenario is unlikely, and \cite{2018arXiv180800052B} showed that it is incompatible with the observations.  The planet is spiraling inward. For a plausible stellar tidal quality factor, the rate of decay is roughly three orders of magnitude too large to be explained by equilibrium stellar tides \citep{2018arXiv180800052B}.  \cite{2017ApJ...849L..11W} showed that dynamical tides can produce a correct decay timescale if the star has evolved off the main sequence. However, recent stellar models by \cite{2018arXiv180800052B} favor a main sequence star rather than a subgiant. 

WASP-12b's orbital decay thus presents a puzzle. We propose that obliquity tides may provide the solution.  When a planet has a non-zero angle between its orbital and spin axes (``obliquity''), tides raised in the planet by the host star produce extremely efficient dissipation \citep[e.g.][]{2007A&A...462L...5L}. A large obliquity can only be maintained in the face of such dissipation if there is an additional torque, for example from the oblate host star or a third body. In these cases a spin-orbit resonance can develop in which the orbital precession is equal to the planet's spin-axis precession. In an upcoming paper  \citep{MillhollandLaughlin2018}, we show that obliquity-driven dissipation may be a key process in sculpting the observed period distribution of the multiple-transiting planetary systems.

\begin{figure*}
\epsscale{1.25}
\plotone{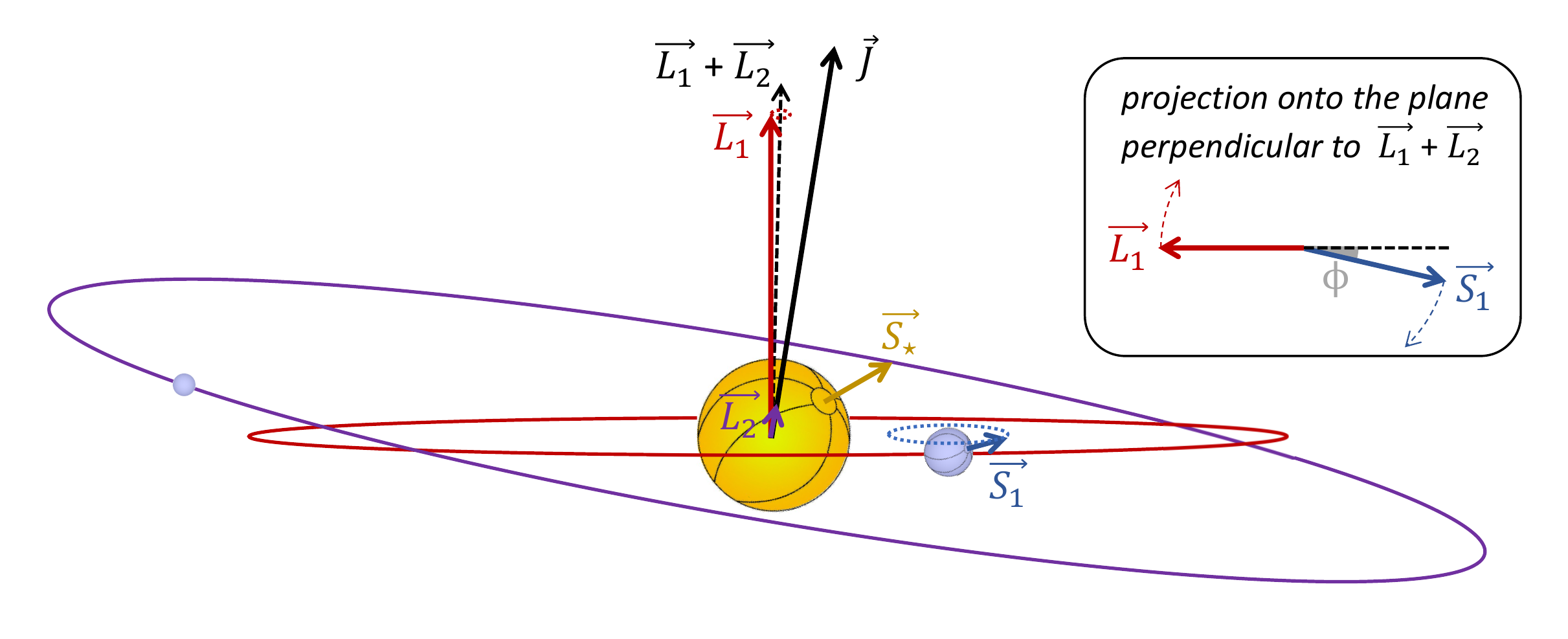}
\caption{Schematic representation of our proposed scenario. Planet b's obliquity maintains a large value as it is forced by a spin-orbit resonance with an exterior, small-mass planet. The angular momentum vectors are drawn roughly to scale, with the exception of $\mathbf{S_1}$. On short timescales, the orbital angular momentum vector, $\mathbf{L_1}$, precesses about $\mathbf{L_1} + \mathbf{L_2}$. The spin angular momentum, $\mathbf{S_1}$, precesses at the same rate, but with a constant phase shift resulting from the tidal dissipation in planet b's interior. (See upper right inset.) On longer timescales, $\mathbf{L_1}$ and $\mathbf{S_{\star}}$ precess around the total vector, $\mathbf{J}$.} 
\label{geometric_diagram}
\end{figure*}

\cite{2005ApJ...628L.159W} suggested that the heat source required to explain the inflated radii of hot Jupiters may, in some cases, be caused by obliquity tides.
\cite{2007A&A...462L...5L}, \cite{2007ApJ...665..754F}, and \cite{2008ASPC..398..281P} later showed this is unworkable if stellar oblateness is the sole driver of orbital precession. The Winn-Holman scenario requires a nearly $90^{\circ}$ obliquity, and the torque induced by the stellar oblateness is too weak to overcome the dissipative tidal torque that acts to damp the obliquity. 
That is, all \textit{isolated} hot Jupiters should have zero obliquities. \cite{2007ApJ...665..754F} investigated whether an exterior perturbing planet could drive the requisite orbital precession. They showed that this is unlikely to induce a high-obliquity state for HD 209458b, but the mechanism has not been thoroughly studied for a generalized hot Jupiter system. Though somewhat fine-tuned, here we show that it \textit{can} operate in the WASP-12 system, potentially generating the observed orbital decay and implying the existence of a readily detectable companion planet. Figure \ref{geometric_diagram} displays a schematic of the proposed set-up. 

 Constraints on WASP-12b's obliquity and tidal quality factor are discussed in Section \ref{section2}. Section \ref{section3} explores the plausible parameters of the perturbing planet. Using these results, Section \ref{section4} presents an N-body simulation in which WASP-12b dissipates due to obliquity tides at a rate matching the observations. Implications and further constraints are discussed in Section \ref{section5}. 

Throughout this work, we adopt the following system parameters. From \cite{2013A&A...551A.108M}, we take $P=1.09142$ days, $R_{\star}={1.63}R_{\odot}$, $\rho_{\star}=0.315\rho_{\odot}$ (which yields $M_{\star}={1.36}M_{\odot}$), $i=83^{\circ}$, and $R_p={1.89}R_{\mathrm{Jup}}$. Using the $K=226$ m/s radial-velocity semi-amplitude obtained by \cite{2009ApJ...693.1920H}, the planet mass is $M_p={1.41}M_{\mathrm{Jup}}$ and $a=0.02299\,\mathrm{AU}$. We take the rotation period of the star to be $P_{\star}=36\,\mathrm{days}$ \citep{2010MNRAS.405.2037W}. Finally, we use the observed period evolution, $P/\dot{P}=-3.2$ Myr, from \cite{2017AJ....154....4P}, so that $a/\dot{a}=-4.8$ Myr.

\section{Obliquity tides}
\label{section2}
Tidal torques act to synchronize a planet's spin, $\omega_p$, and dampen its eccentricity and obliquity, $\epsilon_p$, to zero, such that energy dissipation ceases. Dissipation can continue, however, if non-zero eccentricities or obliquities are maintained by an external driver. The rate at which orbital energy is converted to heat via tides is a strongly increasing function of obliquity.

We begin by calculating the obliquity and tidal quality factor necessary to explain WASP-12b's observed rate of orbital decay. We assume that $e\approx0$ \citep{2011ApJ...727..125C,2011AJ....141...30C,2011MNRAS.413.2500H} and that the spin rate of planet ``b'' is at equilibrium. Using traditional equilibrium tide theory in the viscous approximation \citep{1981A&A....99..126H,1998ApJ...499..853E}, as we will throughout this paper, this rate is \citep{2007A&A...462L...5L}
\begin{equation}
\omega_{p,\mathrm{eq}}=n\frac{2\cos\epsilon_p}{1+\cos^2\epsilon_p}.
\label{omega_eq}
\end{equation}
In the presence of both planetary and stellar equilibrium tides, the secular evolution of the semi-major axis is \citep{2010A&A...516A..64L} 
\begin{equation}
\label{simplified_adot}
\dot{a}=\frac{4a^2}{{G}M_{\star}M_p}\left[K_p\left(\frac{\cos^2\epsilon_p-1}{\cos^2\epsilon_p+1}\right)+K_{\star}\left(\frac{P\cos\epsilon_{\star}-P_{\star}}{P_{\star}}\right)\right].
\end{equation}
$K_p$ is given by 
\begin{equation}
\label{Kp}
K_p=\frac{9}{4Q_p^{\prime}}\left(\frac{G{M_{\star}}^2}{R_p}\right)\left(\frac{R_p}{a}\right)^6n,
\end{equation}
where $Q_p^{\prime}=3Q_p/2k_{2,p}$ is the reduced annual tidal quality factor, with $k_{2,p}$ the Love number. $R_p$ is the planetary radius, $a$ the semi-major axis, and $n=2\pi/P$ the mean-motion.
$K_{\star}$ is defined as in equation \ref{Kp} with $p$ and $\star$ subscripts reversed. 

If $\epsilon_p$ is large, the terms in parentheses multiplying $K_p$ and $K_{\star}$ are both of order unity. We therefore compare the relative strengths of planetary and stellar tides by taking the ratio,
\begin{equation}
\label{K_p_K_star_ratio}
\frac{K_p}{K_{\star}}=\frac{{Q_{\star}}^{\prime}}{Q_p^{\prime}}\left(\frac{M_{\star}}{M_p}\right)^2\left(\frac{R_p}{R_{\star}}\right)^5.
\end{equation}
For plausible tidal quality factors, $Q^{\prime}_{\star}=10^8$ \citep{2018AJ....155..165P,2018MNRAS.476.2542C} and $Q_p^{\prime}\sim10^7$ \citep{2017A&A...602A.107B}, the ratio is $K_p/{K_{\star}}\sim500$. \cite{2018arXiv180800052B} found $a/{\dot{a}}\approx-1.8\,\mathrm{Gyr}$ for equilibrium stellar tides alone. The $\sim 2-3$ orders of magnitude increase from obliquity tides is therefore sufficient to reach the observed rate of orbital evolution, $a/{\dot{a}}=-4.8\,\mathrm{Myr}$.

Ignoring the $K_{\star}$ term in equation \ref{simplified_adot} and solving for the unknowns, $Q_p^{\prime}$ and $\epsilon_p$, yields
\begin{equation}
\label{Obliquity_vs_Qprime_equation}
Q_p^{\prime}\left(\frac{\cos^2\epsilon_p+1}{\cos^2\epsilon_p-1}\right)=\frac{9a^{-11/2}}{\dot{a}}\sqrt{GM_{\star}}\frac{M_{\star}}{M_p}{R_p}^5.
\end{equation}
All quantities on the right hand side are well-determined observationally. 
With the parameter values outlined in the introduction, equation \ref{Obliquity_vs_Qprime_equation} becomes
\begin{equation}
\label{Obliquity_vs_Qprime_constraints}
Q_p^{\prime}\left(\frac{\cos^2\epsilon_p+1}{\cos^2\epsilon_p-1}\right)=-8.59\times10^6.
\end{equation}

\begin{figure}
\epsscale{1.25}
\plotone{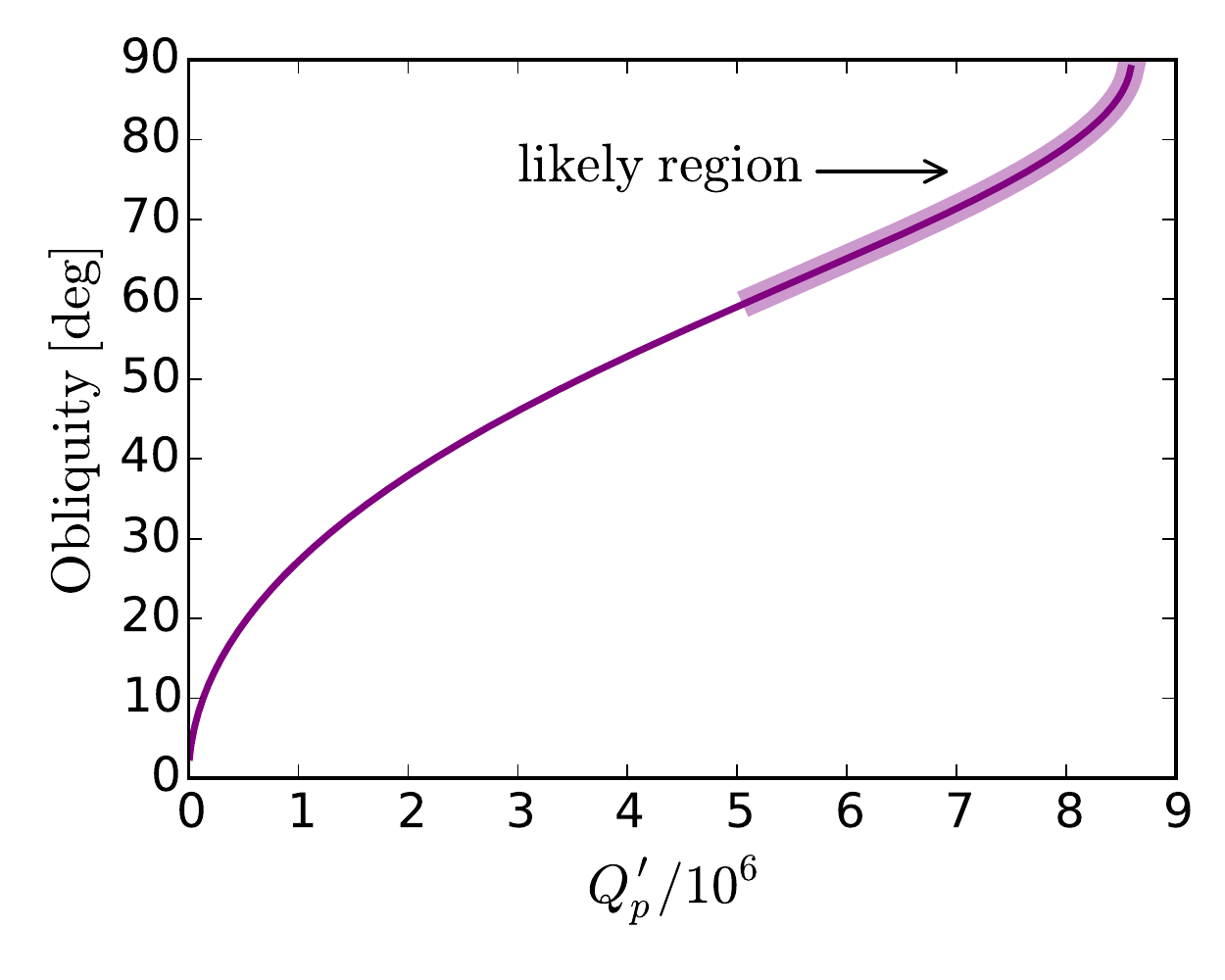}
\caption{Constraints on the obliquity of planet ``b'' and $Q_p^{\prime}=3Q_p/2k_{2,p}$ in order for agreement with the observed rate of orbital decay. The highlighted region is most likely based on the large expected obliquity (see Section \ref{section3}).} 
\label{Obliquity_vs_Qprime}
\end{figure}

Figure \ref{Obliquity_vs_Qprime} shows the resulting mutual constraints on $\epsilon_p$ and $Q_p^{\prime}$. Obliquities, $\epsilon_p\gtrsim30^{\circ}$, are consistent with $Q_p^{\prime}\sim10^6-10^7$. The entirely independent match between the narrow range of values of $Q_p^{\prime}$ required for the obliquity-tide mechanism to work and the \textit{a-priori} expectation for $Q_p^{\prime}$ is an encouraging sign.
In the absence of support, however, tidal torques would damp $\epsilon_p$ to zero in a mere $10^4$ years.
 Large obliquities can be sustained if the planet is locked in a secular spin-orbit resonance that is maintained by another applied torque.

\section{Secular spin-orbit resonance with an exterior perturber}
\label{section3}

In the constant-obliquity configuration, called a ``Cassini state'' \citep{Colombo1966,Peale1969,Ward1975}, the planet's spin and orbital axes precess at the same rate about the same axis, and they are coplanar in the limit of vanishing dissipation.\footnote{There are four Cassini states, but only two of them (states 1 and 2) are stable against tidal dissipation. We refer to Cassini state 2, which is most favorable for maintaining a large obliquity \citep{2007ApJ...665..754F}.} 
We argue that the required orbital precession can arise from secular interactions with an additional planet. If this hypothesis is correct, it places strong constraints on the characteristics of the as-yet undetected perturber.

The torque from the host star on a rotationally-flattened planet will cause the spin-axis to precess about the orbit normal at a period, $T_{\alpha}=2\pi/(\alpha\cos\epsilon_p)$, where $\alpha$ is the precession constant. In the absence of satellites, $\alpha$ is given by \citep{Ward2004,Ragozzine2009}  
\begin{equation}
\alpha=\frac{1}{2}\frac{M_{\star}}{M_p}\left(\frac{R_p}{a}\right)^3\frac{k_{2,p}}{C_p}\omega_p.
\label{alpha}
\end{equation}
$C_p$ is the planet's moment of inertia normalized by $M_p {R_p}^2$. 

The spin-axis precession frequency must be commensurable with the nodal recession frequency, $g=\dot{\Omega}$. In the case that this nodal recession is due to secular perturbations with an exterior planet, the frequency is given by Laplace-Lagrange theory for planets not near mean-motion resonance \citep{1999ssd..book.....M}. To first order in masses and second order in eccentricities and inclinations\footnote{Although the system may not have a small mutual inclination, this expansion is sufficient for a plausibility argument.},  
\begin{equation}
g=-\frac{1}{4}b_{3/2}^{(1)}(\alpha_{12})\alpha_{12}\left(n_1\frac{M_{p2}}{M_{\star}+M_{p1}}\alpha_{12}+n_2\frac{M_{p1}}{M_{\star}+M_{p2}}\right).
\label{g_LL}
\end{equation} 
Here, $\alpha_{12}=a_1/a_2$ and $n_i$ is the mean-motion of planet $i$. We use the subscripts 1 and 2 to refer to planet b and hypothetical planet c, respectively. The constant, $b_{3/2}^{(1)}(\alpha_{12})$, is a Laplace coefficient, defined by 
\begin{equation}
b_{3/2}^{(1)}(\alpha_{12})=\frac{1}{\pi}\int_{0}^{2\pi}\frac{\cos\psi}{(1-2\alpha_{12}\cos\psi+{\alpha_{12}}^2)^{3/2}}d\psi.
\end{equation}

Cassini states obey the resonance condition
\begin{equation}
g\sin(\epsilon_p-I)+\alpha\cos\epsilon_p\sin\epsilon_p=0,
\end{equation}
where $I$ is the inclination of the planet's orbital plane with respect to the invariable plane. If the obliquity is large and the perturbing planet is small, then $I\ll\epsilon_p$ and
\begin{equation}
\lvert{g}\rvert\approx\alpha\cos\epsilon_p.
\label{resonance_condition}
\end{equation}
This condition can be applied to calculate the parameter space in the perturber's mass, $M_{p2}$, and semi-major axis, $a_2$, that allows for commensurability. When WASP-12b is spinning at its equilibrium rate, equation \ref{resonance_condition} becomes 
\begin{equation}
\lvert{g}\rvert=\alpha_{\mathrm{syn}}\frac{2\cos^2\epsilon_p}{1+\cos^2\epsilon_p},
\end{equation}
where $\alpha_{\mathrm{syn}}=\alpha(n/\omega_p)$ is the value of $\alpha$ in the case of synchronous rotation ($\omega_p=n$). The solution for $\epsilon_p$ is 
\begin{equation}
\label{obliquity_at_resonance}
\cos\epsilon_p=\left(\frac{1}{2\alpha_{\mathrm{syn}}/{\lvert{g}\rvert}-1}\right)^{1/2}.
\end{equation}

\begin{figure}
\epsscale{1.25}
\plotone{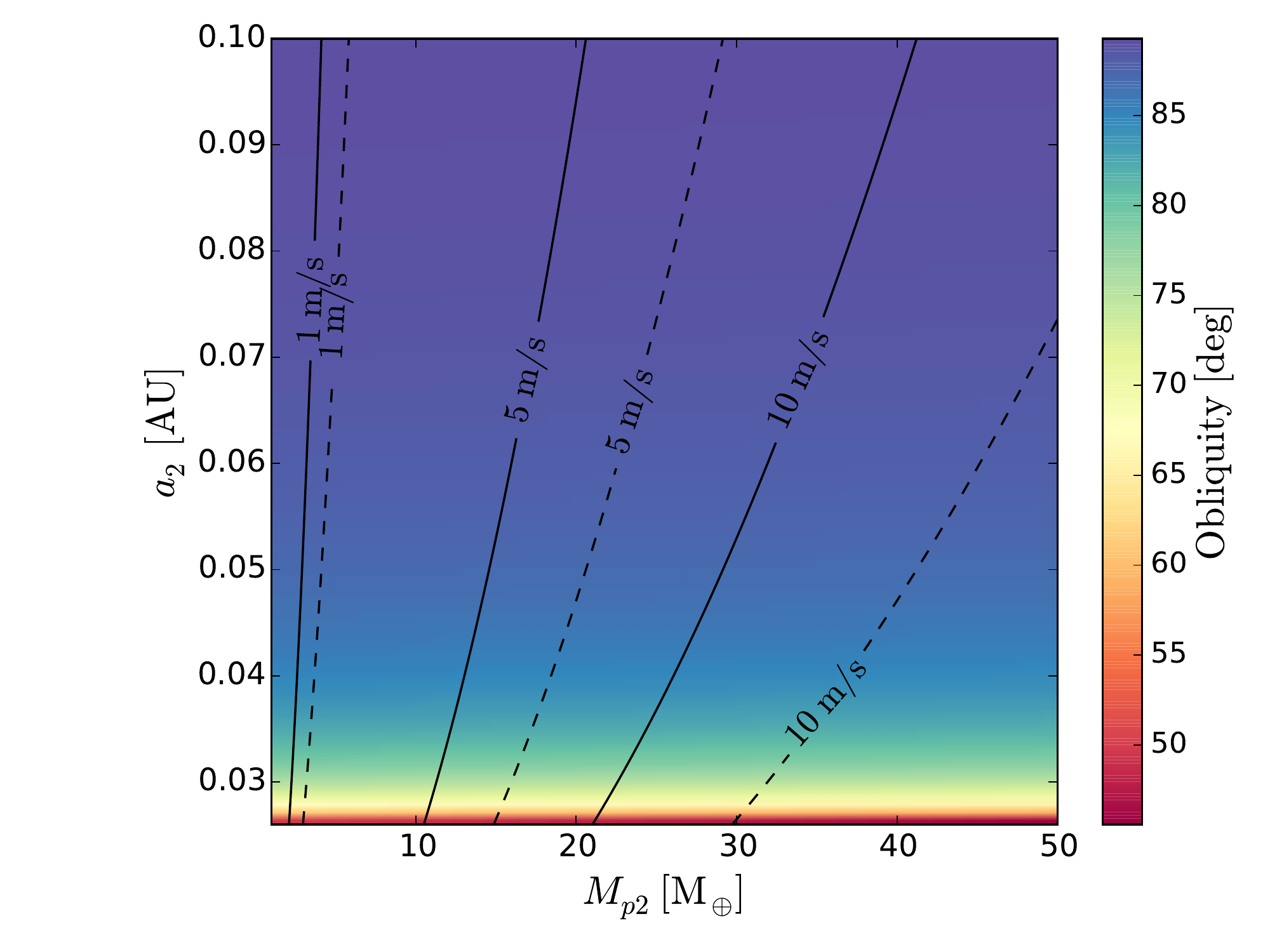}
\caption{A map of the obliquity that WASP-12b would have if it was captured in spin-orbit resonance with a planet of mass, $M_{p2}$, and semi-major axis, $a_2$. The solid/dashed lines are contours of the perturber's RV semi-amplitude for $i_2=90^{\circ}$/$i_2=45^{\circ}$, respectively. 
} 
\label{Obliquity_heatmap}
\end{figure}

In Figure \ref{Obliquity_heatmap}, we show the obliquity necessary for the resonance to hold as a function of $M_{p2}$ and $a_2$. In addition to the parameters outlined in the introduction, we also adopted $k_{2,p}=0.1$ and $C_p=0.2$. The results have little sensitivity to these choices. It is interesting to note that the obliquity is nearly independent of $M_{p2}$. This is because the limit $M_{p2}\ll{M_{p1}}$ makes $g$ only very weakly dependent on $M_{p2}$. 

Overlaid on Figure \ref{Obliquity_heatmap} are contours of the exterior planet's radial velocity (RV) semi-amplitude. 
After subtracting the planet b signal, the standard deviation of the residuals of the published RVs \citep{2009ApJ...693.1920H,2011MNRAS.413.2500H,2014ApJ...785..126K,2017A&A...602A.107B} is $\sim16\,\mathrm{m/s}$. The RV semi-amplitude of the hypothetical planet must be less than this.

\subsection{Secular orbital evolution of planet b}
\label{section3.1}

The secular decrease in planet b's semi-major axis induced by tides causes $\alpha$ to increase and $\lvert{g}\rvert$ to decrease.
As a result, the obliquity must increase adiabatically in order to maintain the resonance (equation \ref{resonance_condition}). This in turn increases the rate of tidal dissipation. Orbital decay due to obliquity tides is therefore a runaway process.

Here we examine the timescale of this runaway and estimate planet b's past semi-major axis evolution. We do this semi-analytically by coupling the secular solution for $\dot{a}_1$ (equation \ref{simplified_adot}) with the resonant solution for the obliquity (equation \ref{obliquity_at_resonance}) in an ODE solver. $M_{p2}$ and $a_2$ provide the only sensitive dependencies for $a_1(t)$. It also depends on the unknowns $k_{2,p}$ and $C_p$, which we fix  to the fiducial values noted above.

\begin{figure}
\epsscale{1.25}
\plotone{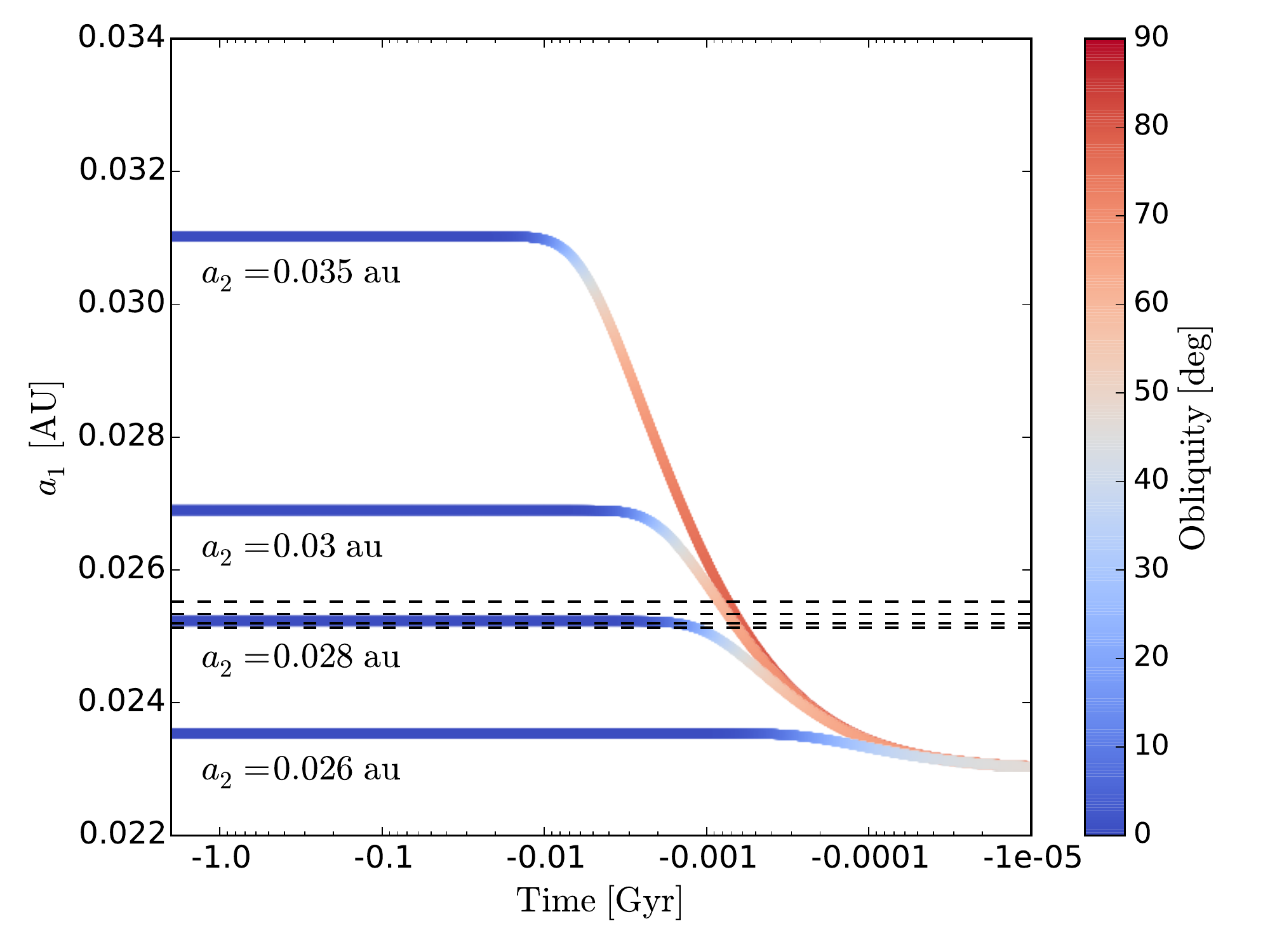}
\caption{Backwards-in-time tidal evolution curves for $a_1$ corresponding to four different $a_2$. Planet b's obliquity is assumed to evolve adiabatically according to maintenance of the spin-orbit resonance (equation \ref{obliquity_at_resonance}). The four horizontal lines show, for each value of $a_2$, the maximum initial values of $a_1$ assuming that angular momentum conservation is preserved by damping mutual inclination.} 
\label{a1_obliquity_evolution}
\end{figure}

Figure \ref{a1_obliquity_evolution} shows four evolutionary trajectories for $a_1$. They all use $M_{p2}=20\,M_{\oplus}$ but have different values of $a_2$. 
The initial obliquities were set to the resonant values given by equation \ref{obliquity_at_resonance}, and $Q_p^{\prime}$ was calculated via equation \ref{Obliquity_vs_Qprime_constraints} so as to recover the observed present-day value of $\dot{a}_1$. The system must conserve total angular momentum as planet b inspirals, which can be accomplished through alignment of $\bm{L_1}$ with $\bm{L_2}$ and/or $\bm{S_{\star}}$  \citep{2007ApJ...665..754F}. If alignment only occurs between $\bm{L_1}$ and $\bm{L_2}$, we calculate the maximum initial value of $a_1$ by assuming the mutual inclination, $\Delta\,i$, is currently small but was initially near $90^{\circ}$ \citep{2016ApJ...829..114B}.  
We note that the increase in $\Delta\,i$ will slightly modify equation \ref{g_LL} for $g$ and the timescale at which $a_1(t)$ evolves. 

There are several features of interest in Figure \ref{a1_obliquity_evolution}. 
First, if angular momentum conservation is preserved solely by aligning $\bm{L_1}$ and $\bm{L_2}$ (and not $\bm{S_{\star}}$), there are strict constraints on the initial value of $a_1$. 
Implications of this are discussed in Section \ref{section5}.
Second, it is possible that the obliquity reaches $\epsilon_p\lesssim1^{\circ}$ within a billion years into the past of the system's $1.7\pm0.8\ \mathrm{Gyr}$ life \citep{2011AJ....141..179C}. This is important because it implies that the resonance could have been captured when $\epsilon_p\lesssim1^{\circ}$, yet $\epsilon_p$ still reached large values by the present day. This makes it unnecessary to invoke an unrealistic large primordial obliquity that might have made the initial resonant capture difficult to explain.

\section{Example simulation}
\label{section4}

Our investigation thus far has established the plausibility that obliquity tides may be acting on WASP-12b. The analysis, however, has so far only been analytic. Numerical simulations can substantiate the hypothesis by confirming the configuration's stability. Here we present an example simulation that is consistent with all constraints outlined above. We adopt the following parameters for the perturbing planet: $a_2=0.04\ \mathrm{AU}$, $M_{p2}=20\ M_{\oplus}$,  $R_{p2}=5\ R_{\oplus}$, $k_{2,p2}=0.3$, $C_{p2}=0.25$, and $\lvert{i_1-i_2}\rvert=20^{\circ}$. We use equations \ref{Obliquity_vs_Qprime_equation} and \ref{obliquity_at_resonance} to determine the value of planet b's $Q_p$ that agrees with a resonant solution and the observed orbital decay.

We model the tidal, spin, and orbital evolution of the WASP-12 system consisting of the host star, planet b, and the additional planet. 
Our code consists of direct numerical integrations using instantaneous accelerations in the framework of \cite{2002ApJ...573..829M}. 
In addition to the standard Newtonian gravitational accelerations, we also apply accelerations on the planets due to (1) the quadrupolar gravitational moment of the star and (2) equilibrium tides raised in the planets from the star \citep{1981A&A....99..126H,1998ApJ...499..853E}. We evolve the orbital and spin equations in hierarchical (Jacobi) coordinates using a Bulirsch-Stoer integrator \citep{1992nrfa.book.....P} with the timestep equal to $0.01P_1$ and the timestep accuracy parameter set to $\eta=10^{-13}$.

A simulation of the present-day system is indifferent to when and how the Cassini state was originally captured. Figure \ref{a1_obliquity_evolution} shows that this capture is straightforward to explain because it could have occurred with $\epsilon_p\lesssim1^{\circ}$. The capture process requires that $T_{\alpha}=2\pi/(\alpha\cos\epsilon_p)$ and $T_g=2\pi/\lvert{g}\rvert$ evolve such that $T_{\alpha}/T_g$ crosses through unity from above. There are many potential scenarios for this (e.g., a decrease in $T_{\alpha}$ due to tidal synchronization of planet b's spin). Here we solely wish to verify that a scenario consistent with the constraints is capable of existing \textit{today}, so we use a contrived mechanism to induce the capture. We start with $\epsilon_p=88^{\circ}$ and $Q_p$ ten times larger than the target value. The obliquity damps due to the tidal torque, and $T_{\alpha}/T_g$ slowly crosses unity from above. After the resonant locking, we decrease $Q_p$ on a $10,000$ year exponential timescale until it reaches the target value.

\begin{figure}
\epsscale{1.25}
\plotone{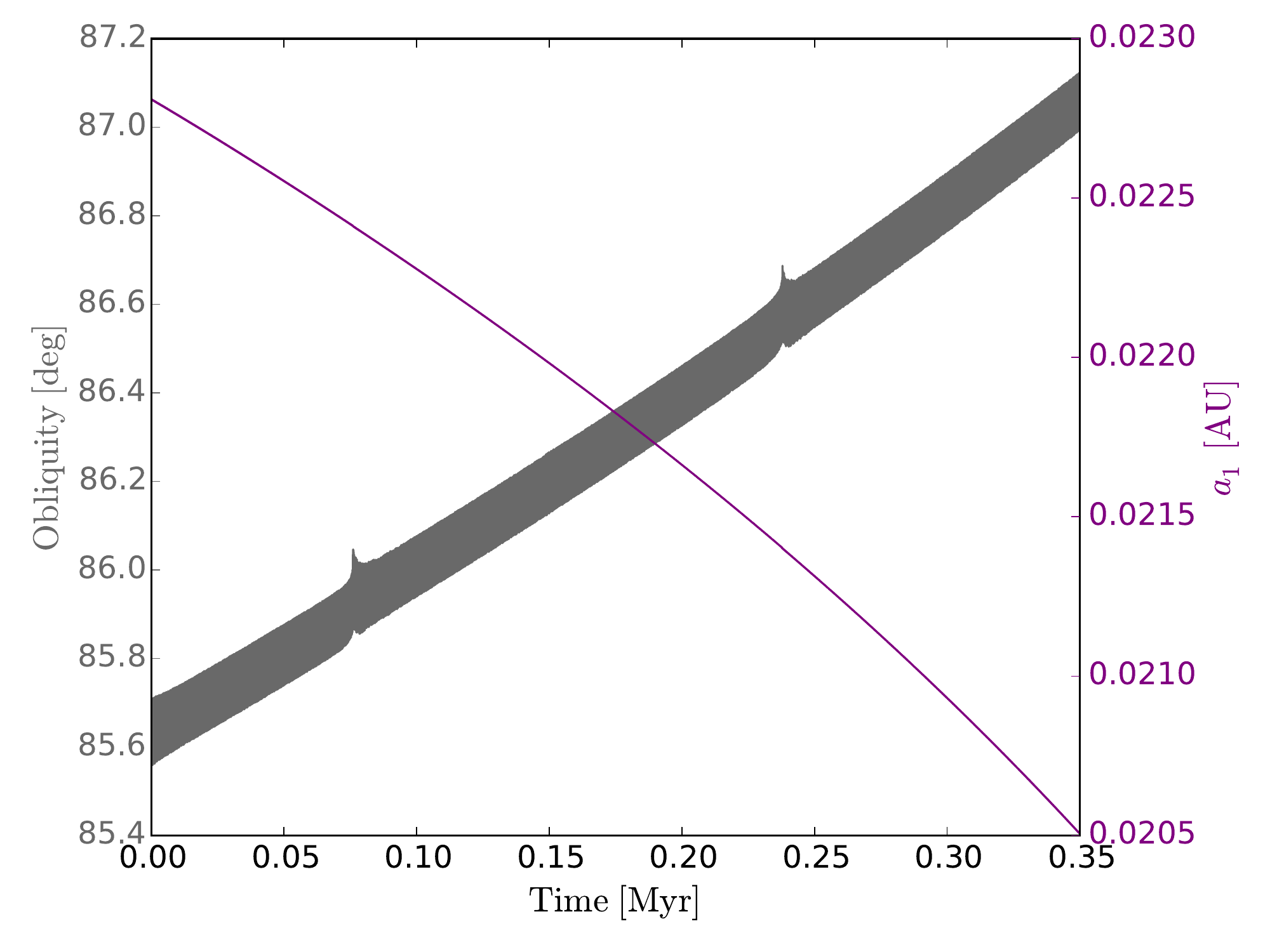}
\caption{The evolution of planet b's obliquity (gray) and semi-major axis (purple) in the example simulation. The jumps near $\sim0.075\ \mathrm{Myr}$ and $\sim0.24\ \mathrm{Myr}$ are due to encounters with the 7:3 and 5:2 mean-motion resonances resulting from the divergent tidal migration.} 
\label{planet_b_obliquity_evolution}
\end{figure}

Figure \ref{planet_b_obliquity_evolution} shows the time evolution of planet b's obliquity and semi-major axis during a period well after the capture has taken place and $Q_p$ has been reduced to the target value. At the beginning of the simulation, the semi-major axis is evolving at a rate, $\dot{a}_1\approx-0.0048\ \mathrm{AU/Myr}$, in agreement with current observations. $\lvert\dot{a}_1\rvert$ increases as time advances, and $\epsilon_p$ increases adiabatically so as to maintain the resonance. 

For a dissipative Cassini state,
there exists an upper limit at which the tidal torque is as strong as the perturbation torque (here due to the additional planet); beyond that, the resonance can no longer be maintained \citep{2007ApJ...665..754F,2008ASPC..398..281P}. 
In our example simulation, the spin vector is phase shifted by $\phi=13^{\circ}$ out of the plane defined by $\mathbf{L_1}$ and $\mathbf{J}$. \cite{2007ApJ...665..754F} showed that the dissipative limit is at $\phi=90^{\circ}$, so $13^{\circ}$ is well away from this. Although the phase shift increases as $a_1$ decreases and the Cassini state will eventually break, the simulation indicates that the present-day WASP-12 system can most certainly exist within the dissipative limit.

\section{Discussion}

\label{section5}
WASP-12b has been the subject of intense investigation since its discovery in 2009. The obliquity tide hypothesis that we have put forth to explain the planet's rapid orbital decay can resolve additional mysteries. The planet's thermal phase curve presents unusual features \citep{2012ApJ...747...82C,2018AJ....156...28A} that might be naturally explained using a nonzero obliquity model \citep{AdamsMillhollandLaughlin2018}. Secondly, despite its scorching temperature ($T_{\mathrm{eq}}\approx2500\ \mathrm{K}$), the planet's extreme radius inflation is considered an outlier that cannot be explained without an extra heating mechanism \citep{2009ApJ...702.1413M,2011ApJ...727...75I,2011ApJS..197....9F}. Obliquity tides can readily provide the heat. If we assume that the energy from orbital decay is radiating from the planet, the energy involved is
\begin{equation}
\frac{dE}{dt}=\frac{(GM_{\star})^{3/2}M_{p}}{6\pi}a^{-5/2}\dot{P}\simeq5\times10^{30}\,{\rm erg\,s^{-1}}
\end{equation}
implying a tidal luminosity that is of the same order as the insolation received by the planet.

Extreme tidal heating is consistent with the observation that WASP-12b is overflowing its Roche lobe \citep{2012ApJ...760...79H}
and losing mass at an impressive rate of $\sim1.6\ M_{\oplus}/{\mathrm{Myr}}$ \citep{2017ApJ...835..145J}. These independent signs that the planet is undergoing an end-of-life thermal runaway are compatible with the obliquity tide hypothesis. 

To explain the peculiarities of WASP-12, we have invoked a peculiar configuration: a hot Jupiter with an inclined, small planetary companion on a close-in orbit. We investigated the alternative that the requisite orbital precession could be supplied by the torque from stellar oblateness rather than a companion planet \citep{2007ApJ...665..754F}. However, this results in $\lvert g \rvert$ so small that $\epsilon_p$ is nearly $90^{\circ}$, and dissipative torque balance is impossible. Moreover, in this scenario, $\epsilon_p$ would decrease as the orbit decays, and this is inconsistent. 

While no such difficult-to-detect \citep{2016ApJ...823L...7M,2017AJ....154...83M}, misaligned system is currently known to exist, they are the primary prediction of \cite{2016ApJ...829..114B}'s theory of \textit{in situ} formation of hot Jupiters (see also \citealt{2017AJ....154...93S}). 
Confirmation of our obliquity tide hypothesis would make WASP-12 the second system definitively known to host both a hot Jupiter and a nearby companion \citep{2015ApJ...812L..18B} and would provide evidence in support of \cite{2016ApJ...829..114B}'s prediction.

Further investigations may strengthen or weaken the obliquity tide hypothesis. First, as noted in section \ref{section3.1}, the system must conserve angular momentum and if this is solely via realignment of $\bm{L_1}$ and $\bm{L_2}$, then $a_2\lesssim0.027\,\mathrm{AU}$ is preferred. Though this renders the obliquity tide hypothesis uncomfortably fine-tuned, the rarity of planet b's observed orbital decay favors a short lifetime of the configuration and therefore a small $a_2$.
There is, however, a significant storage of misaligned ($\lambda ={59^{+15}_{-20}}^{\circ}$; \citealt{2012ApJ...757...18A}) angular momentum in $\bm{S_{\star}}$. If there is a mechanism for aligning $\bm{L_1}$ and  $\bm{S_{\star}}$, the constraints are much less stringent; the limit on the initial value of $a_1$ is $\sim0.043\,\mathrm{AU}$ using a conservative estimate of $P_{\star}$. 

Another test of the hypothesis arises from Transit Duration Variations (TDVs) of planet b due to its orbital precession \citep{2002ApJ...564.1019M}. The precession period is $\sim15\,\mathrm{years}$,
and for mutual orbital inclinations, $\Delta{i}\lesssim 20^{\circ}$, we calculate that the \textit{maximum} peak-to-peak TDV amplitude is
\begin{equation}
\Delta{T_{\mathrm{dur}}}\approx0.184\,\mathrm{min}\left(\frac{\Delta{i}}{1^{\circ}}\right)\left(\frac{M_{p2}}{10{M_{\oplus}}}\right).
\end{equation}
If $\Delta{i}\lesssim15^{\circ}$, TDVs would not have been detected in the extant photometry, which has $\sim 3$ min transit duration uncertainties \citep{2017AJ....153...78C}. 
In addition to TDVs, the perturbing planet may also induce Transit Timing Variations (TTVs). \cite{2017AJ....153...78C} ruled out \textit{sinusoidal} TTVs with amplitude $\gtrsim35\,\mathrm{sec}$. If the perturbing planet is non-resonant, the TTVs would certainly be smaller than this. Future photometric monitoring (e.g. by \textit{TESS}) will test our hypothesis by placing tighter constraints on TTVs/TDVs. 

RV follow-up is likely the best avenue for falsification of our hypothesis. The RV semi-amplitude of the perturbing planet in our example simulation was $K\sim7$m/s, and planets with $K\lesssim10$m/s would not yet have been detected in the current dataset. This makes WASP-12 an excellent candidate for future high-precision RV surveys.

\section{Acknowledgements}
We thank the anonymous referee, whose insightful review improved the quality of this work. S.M. is supported by the NSF Graduate Research Fellowship Program under Grant  DGE-1122492. G.L. acknowledges NASA Astrobiology Institute support under Agreement \#NNH13ZDA017C.


\end{document}